\documentclass[sigconf,balance=true,hidelinks,table]{acmart}

\usepackage{hyperref}
\usepackage{amsmath,amsfonts}
\usepackage{algorithmic}
\usepackage[ruled,linesnumbered,noend]{algorithm2e}
\usepackage{xcolor}
\usepackage{svg}
\usepackage{graphicx}
\usepackage{textcomp}
\usepackage{xcolor}
\usepackage{nccmath}
\usepackage{environ}
\usepackage{tcolorbox}
\usepackage{soul}
\usepackage{booktabs}
\usepackage{multirow}
\usepackage{array}
\usepackage{float}
\usepackage{makecell}
\usepackage{flushend}
\usepackage{enumitem}
\usepackage{hhline}
\usepackage{comment}
\usepackage{epstopdf} 
\usepackage{caption}
\usepackage{subcaption}

\usepackage{xcolor}
\usepackage{listings}

\usepackage{scalerel}

\NewEnviron{myequation}{%
\begin{equation}
\scalebox{1}{$\BODY$}
\end{equation}
}

\newcommand{\fig}{Fig.}
\newcommand{\tab}{Table}

\newcolumntype{P}[1]{>{\centering\arraybackslash}p{#1}}
\newcolumntype{R}[1]{>{\raggedleft\arraybackslash}p{#1}}

\newcommand{\code}[1]{\textnormal{\texttt{#1}}}

\usepackage{tikz}
\usetikzlibrary{shapes.arrows}

\definecolor{brightgreen}{rgb}{0.4, 1.0, 0.0}
\definecolor{ferrarired}{rgb}{1.0, 0.11, 0.0}

\acmConference[ESEM 2024]{The 18th ACM/IEEE International Symposium on Empirical Software Engineering and Measurement}{20–25 October, 2024}{Barcelona, Spain}

\setcopyright{none}

\begin{document}

\title{Mitigating Data Imbalance for Software Vulnerability Assessment:\\Does Data Augmentation Help?}

\author{Triet Huynh Minh Le}
\affiliation{\institution{CREST - The Centre for Research on Engineering Software Technologies, The University of Adelaide}
\city{Adelaide}
\country{Australia}}
\affiliation{\institution{Cyber Security Cooperative Research Centre, Australia}
\city{}
\country{}}
\email{triet.h.le@adelaide.edu.au}

\author{M. Ali Babar}
\affiliation{\institution{CREST - The Centre for Research on Engineering Software Technologies, The University of Adelaide}
\city{Adelaide}
\country{Australia}}
\affiliation{\institution{Cyber Security Cooperative Research Centre, Australia}
\city{}
\country{}}
\email{ali.babar@adelaide.edu.au}

\begin{abstract}

\textbf{Background}: Software Vulnerability (SV) assessment is increasingly adopted to address the ever-increasing volume and complexity of SVs.
Data-driven approaches have been widely used to automate SV assessment tasks, particularly the prediction of the Common Vulnerability Scoring System (CVSS) metrics such as exploitability, impact, and severity. SV assessment suffers from the imbalanced distributions of the CVSS classes, but such data imbalance has been hardly understood and addressed in the literature.
\textbf{Aims}: We conduct a large-scale study to quantify the impacts of data imbalance and mitigate the issue for SV assessment through the use of data augmentation.
\textbf{Method}: We leverage nine data augmentation techniques to balance the class distributions of the CVSS metrics. We then compare the performance of SV assessment models with and without leveraging the augmented data.
\textbf{Results}: Through extensive experiments on 180k+ real-world SVs, we show that mitigating data imbalance can significantly improve the predictive performance of models for all the CVSS tasks, by up to 31.8\% in Matthews Correlation Coefficient.
We also discover that simple text augmentation like combining random text insertion, deletion, and replacement can outperform the baseline across the board.
\textbf{Conclusions}: Our study provides the motivation and the first promising step toward tackling data imbalance for effective SV assessment.

\end{abstract}


\maketitle

\section{Introduction}

Software Vulnerabilities (SVs) like Heartbleed~\cite{heartbleed} or Log4Shell~\cite{log4j} are security bugs that adversely affect the quality of software systems, potentially leading to catastrophic cybersecurity attacks~\cite{ghaffarian2017software}.
In practice, fixing all detected SVs simultaneously is not always feasible due to time and resource constraints~\cite{khan2018review,le2021large}.
Rather, a more practical approach involves prioritizing SVs posing serious and impending security threats, which usually requires inputs from SV assessment~\cite{smyth2017software,le2022survey,le2022towards}.
SV assessment identifies diverse attributes such as exploitability, impact, and severity levels of SVs~\cite{kamongi2013vulcan}. For instance, SVs with a substantial likelihood of exploitation and severe consequences typically demand elevated priority for resolution.
Currently, Common Vulnerability Scoring System~\cite{cvss} (CVSS) is the most commonly used industry-grade standard for SV assessment. However, the assignments of these CVSS metrics to ever-increasing SVs are tedious and time-consuming for security experts~\cite{feutrill2018effect}, which has motivated the research on automated approaches for the tasks.

An increasing number of studies have proposed various data-driven techniques to automate the prediction of the CVSS metrics by leveraging available SV data in the wild (e.g.,~\cite{yamamoto2015text,le2019automated,spanos2018multi,han2017learning,gong2019joint}).
Most of these prediction models automatically learn the patterns from textual SV descriptions published on SV repositories/databases such as National Vulnerability Database (NVD)~\cite{nvd_website} to classify the CVSS metrics.
The current literature has explored different modeling algorithms ranging from traditional Machine Learning (ML) techniques to more advanced Deep Learning (DL) architectures to perform the classifications~\cite{le2020deep,le2022survey,arani2024systematic}.
However, the development and quality of these classification models could be negatively affected by the \textit{data imbalance} issue; i.e., a data class has significantly fewer samples than the other classes~\cite{kaur2019systematic}.

We argue that data imbalance does exist in CVSS-based SV assessment, but it is being overlooked by the current literature.
Our analysis of the SVs published on NVD from 1988 to 2023 showed that all the CVSS metrics used for SV assessment, on average, had a (minority) class constituting only 10.2\% of total samples, being around six times smaller than that (61.1\%) of the respective majority class.
The data imbalance issue has been shown to significantly impede the performance of downstream prediction models in various classification tasks (e.g.,~\cite{malhotra2017empirical,rodriguez2014preliminary,song2018comprehensive,liu2022comparative}).
Although our aforementioned analysis has clearly shown the presence of data imbalance in CVSS-based SV assessment, little is known about the potential impact/benefits of mitigating the issue for the tasks.

Data augmentation has been one of the most widely used tools to gauge the impact and provide mitigation of the data imbalance issue in the ML/DL domains~\cite{feng2021survey}.
This approach aims to artificially increase the number of samples, which can adjust the class distributions of data during model training, making the model less biased towards the majority class(es).
Given that SV assessment is currently using textual SV descriptions as input, existing data augmentation techniques for text data would be, in principle, applicable to this domain.
It is worth noting that these data augmentation techniques have been demonstrated to be effective for various text classification tasks~\cite{bayer2022survey}.
Nevertheless, to the best of our knowledge, there has been no systematic investigation into the extent to which these data augmentation techniques are useful for tackling the data imbalance issue for SV assessment.

To bridge this gap, we aim to conduct the first large-scale study on the potential utilization of data augmentation for quantifying and mitigating the data imbalance issue in the development of SV assessment models.
We first investigate nine different data augmentation techniques to generate augmented SV descriptions from the original descriptions of 180,087 real-world SVs. Then, we compare the predictive performance of SV assessment models \textit{with} and \textit{without} using the augmented descriptions.
These models leverage commonly used ML and DL techniques to automate the classification of the seven CVSS metrics, i.e., Access Vector, Access Complexity, Authentication, Confidentiality, Integrity, Availability, and Severity.

Our key \textbf{contributions} can be summarized as follows:

\begin{itemize}
[noitemsep,topsep=0pt,leftmargin=*]
  \item Through the lens of data augmentation, we are the first to systematically investigate the significance and impacts of mitigating the data imbalance issue on the SV assessment models based on SV reports/descriptions.
  Our findings show that addressing data imbalance can improve the performance of SV assessment models by 5.3--31.8\% in Matthews Correlation Coefficient (MCC) across the seven CVSS metrics.
  \item We empirically benchmark the effectiveness of different data augmentation techniques for SV assessment.
  We find that a combination of random text insertion, deletion, and substitution/replacement produces the highest performance averaging all models and tasks, i.e., 11.3\% better MCC than the baseline without using data augmentation.
  We also shed light on the best techniques for individual tasks and models that help achieve new heights in performance for SV assessment.
  \item We share the code and models at~\cite{reproduction_package_esem2024}  to reproduce the results and facilitate future research in this direction.
\end{itemize}

\noindent Overall, our study sheds light on the possibility of using data augmentation to enhance SV assessment. Our findings are expected to provide baselines for researchers to further explore this direction and improve the performance of SV assessment, which in turn can enable more effective SV mitigation/fixing for practitioners.

\noindent \textbf{Paper structure}. Section~\mbox{\ref{sec:background}} introduces SV assessment tasks and the potential of data augmentation for the tasks.
Section~\mbox{\ref{sec:research_methods}} presents the research questions and the respective research methods to answer these questions. Section~\mbox{\ref{sec:results}} shows the experimental results of each question.
Section~\mbox{\ref{sec:discussion}} discusses the findings and the threats to validity.
Section~\ref{sec:related_work} covers the related work.
Section~\mbox{\ref{sec:conclusions}} concludes the study.

\section{Background and Motivation}\label{sec:background}

\subsection{CVSS-Based SV Assessment}
\label{subsec:sv_assessment_bg}

SV assessment, occurring between the detection and remediation phases in the SV management lifecycle, reveals various characteristics of SVs identified earlier~\cite{foreman2019vulnerability}.
In practice, an average of 80 new SVs are discovered daily~\cite{sv_number_nvd}, and each SV may require up to 1.5 hours to fix~\cite{cornell2012remediation,ben2015factors}, totaling 120 hours needed for fixing. These statistics mean that there is certainly not sufficient time to fix all of these SVs within a 24-hour day, so the fixing prioritization of SVs is inevitable.
SV assessment pinpoints ``hot spots'' in terms of security risks, demanding more attention in a system.
Accordingly, the assessment guides the development of an efficient prioritization for SV fixing based on resource/time availability.

Common Vulnerability Scoring System (CVSS)~\cite{cvss} has been widely regarded as the standard framework by both researchers and practitioners for conducting SV assessment.
Despite newer versions of CVSS, version 2 is still the most commonly used because its assessment of old SVs is still relevant.
An illustrative example is the SV with CVE-2004-0113 which was discovered in 2004 and was still exploited in 2018~\mbox{\cite{old_sv_exploit}}.
Therefore, in this study, we choose the metrics of CVSS version 2 as the outputs for our SV assessment models. In this paper, we use the term ``CVSS metrics'' to mainly refer to version 2 of the CVSS metrics unless specified otherwise.

CVSS version 2 offers metrics that gauge three primary facets of SVs: \textit{Exploitability}, \textit{Impact}, and \textit{Severity}. \textbf{Exploitability} evaluates Access Vector (i.e., the medium/technique to penetrate a system), Access Complexity (i.e., the complexity to initiate an attack), and Authentication (i.e., whether/what authentication the attack requires).
Meanwhile, \textbf{Impact} metrics focus on the Confidentiality, Integrity, and Availability attributes of the system of interest in case of exploitation. \textbf{Severity} is then determined based on both Exploitation and Impact metrics, which approximates the criticality of detected SVs.
However, Severity is a high-level combination of Exploitability and Impact, and thus, it does not provide a full understanding of SVs, potentially leading to a sub-optimal SV fixing plan.
For example, according to the CVSS specification~\cite{cvss_v2}, two SVs would have the same severity level if they share the same exploitability but affect different attributes (e.g., Confidentiality vs. Integrity) of a system to the same extent.
As a result, to ensure a thorough assessment of SVs, we utilize all of the seven CVSS version 2 metrics (i.e., Confidentiality, Integrity, Availability, Access Vector, Access Complexity, Authentication, and Severity) as the outputs for building SV assessment, akin to prior studies (e.g.,~\cite{spanos2018multi,le2019automated,gong2019joint,le2021deepcva}).

\begin{figure}[t]
  \centering
  \includegraphics[trim={20cm 3.6cm 26cm 3.3cm},clip,width=\columnwidth,keepaspectratio]{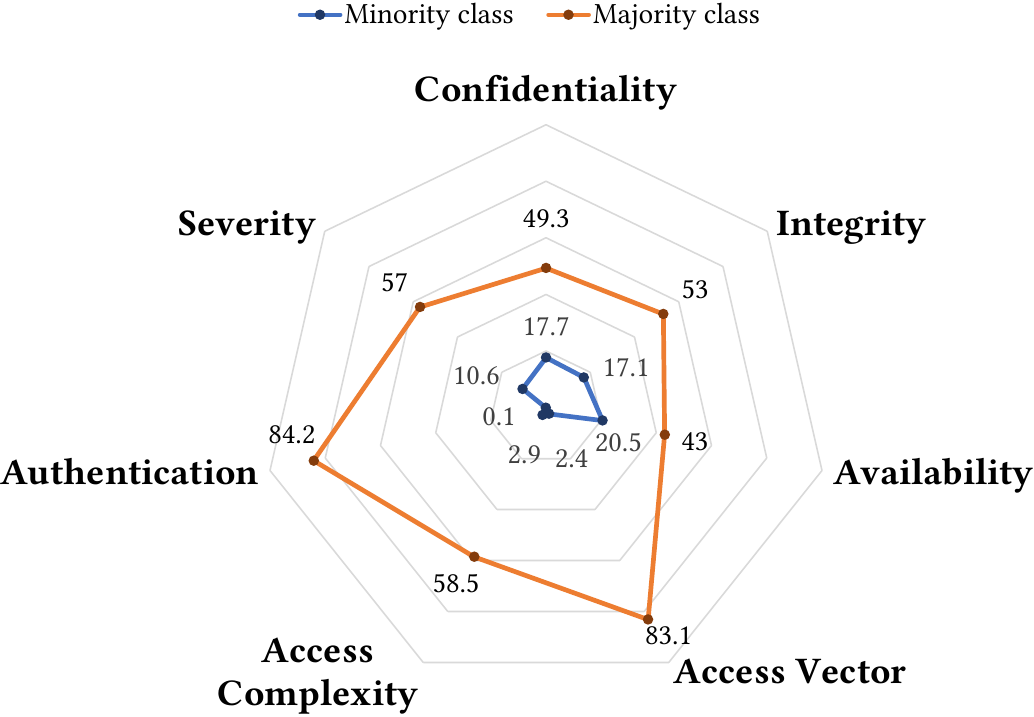}
  \caption{Data percentages (\%) of the minority and the majority classes of the seven CVSS metrics of the SVs collected from National Vulnerability Database, illustrating the data imbalance issue for SV assessment. \textbf{Note}: The percentages do not add up to 100\% as each CVSS metric has three classes.}
  \label{fig:data_imbalance}
\end{figure}

Despite the evident benefits of the CVSS metrics, it is extremely challenging for security experts to manually assign these metrics for ever-increasing SVs. It has been shown that it can take up to 100+ days for the CVSS metrics to be assigned to an SV, mainly due to tedious manual assignment and verification processes~\cite{gong2019joint}. To help alleviate such burden for security experts, many of the current studies have relied on \textit{descriptions} in SV reports to develop Machine Learning (ML)/Deep Learning (DL) based techniques to automatically predict missing CVSS metrics (e.g.,~\cite{han2017learning,spanos2018multi,gong2019joint,le2019automated,elbaz2020fighting}).
These textual descriptions contain various insights about SVs, which can be leveraged for SV assessment. For example, the description of CVE-1999-0315 is ``\textit{Buffer overflow in Solaris fdformat command gives root access to local users}'', distilling the location (i.e., Solaris fdformat command), type (i.e., buffer overflow), and impact/consequence (i.e., giving root access) of the SV.
Moreover, such descriptions are almost always present when new SVs are published. The useful information and availability of SV descriptions have made them valuable inputs/resources for timely CVSS-based SV assessment using data-driven approaches~\cite{le2022survey}.

\begin{figure*}[t]
  \centering
  \includegraphics[width=\textwidth,keepaspectratio]{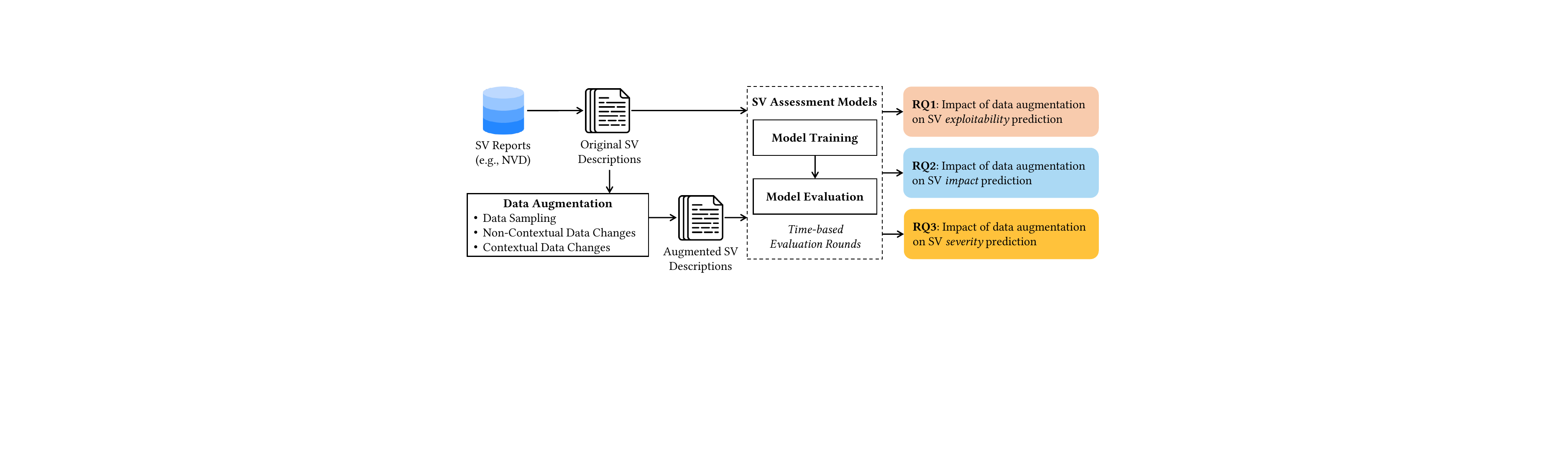}
  \caption{Overview of the research methods used for the investigation of data augmentation for different SV assessment tasks.}
  \label{fig:overview}
\end{figure*}

\subsection{Data Augmentation for SV Assessment with Data Imbalance}
\label{subsec:data_aug_bg}

A key challenge with CVSS-based SV assessment using ML/DL is \textit{data imbalance}, which is illustrated in \fig~\ref{fig:data_imbalance}. Data imbalance occurs when the number of samples of a (minority) class is much smaller than those of the other (majority) classes.
Based on more than 180k SVs we collected from National Vulnerability Database~\cite{nvd_website}, we found that the minority classes existed for all the CVSS metrics.
Notably, many of the minority classes (i.e., Access Vector, Access Complexity, and Authentication) even constituted less than 3\% of the total data.
Moreover, SVs with \textit{Complete} levels of Confidentiality, Integrity, and Availability impacts are of practical importance and require special attention to address, but they were only the minority classes.
Such data imbalance can make assessment models struggle to capture data patterns and provide accurate predictions for these small-sized yet important/relevant classes.
This issue has also been shown to hinder the overall model performance and usefulness in other domains~\cite{kaur2019systematic}.
However, the degree of impact that the data imbalance problem has on SV assessment is still largely unexplored.
The potential revelation of such impact with the use of data augmentation is given hereafter.

Data augmentation involves techniques that artificially generate new/additional data samples to increase the data size~\cite{mumuni2022data}.
Data augmentation can change the class distributions and has been shown to reduce the likelihood of overfitting for models.
The most straightforward way of data augmentation is to simply duplicate existing samples, namely random over-sampling.
Generally, this technique can be applied to any data-driven tasks, but it is yet to be used for SV assessment.
In the context of SV assessment, the input is textual SV descriptions, as described in Section~\ref{subsec:sv_assessment_bg}, which also motivates the exploration of text-based (data) augmentation techniques.
These techniques basically make small changes to the input text or change the text in a way that can still preserve the overall meaning of the input text.\footnote{More details about these techniques are given in Section~\ref{subsec:studied_data_aug}.}
Such augmented data can also improve the model performance in many downstream text classification tasks~\cite{bayer2022survey}.
However, to the best of our knowledge, there has been no study exploring the potential and use of these text-based augmentation for SV assessment tasks.
To bridge this gap, we study the extent to which different data augmentation techniques, including general sampling and text-specific ones, can enhance the model performance and in turn highlight the impact/importance of mitigating the data imbalance issue for the tasks.

\section{Case Study Design and Setup}
\label{sec:research_methods}

We outline the setup for investigating the use of data augmentation techniques for SV assessment. Section~\ref{subsec:rqs} describes the two research questions.
\fig~\ref{fig:overview} depicts the research methods used to answer the questions. Given the benefits mentioned in Section~\ref{subsec:data_driven_sap}, SV descriptions collected from SV reports on NVD were used as the main input in our study; their details are given in Section~\ref{subsec:dataset}. Such SV descriptions were then used by the data augmentation techniques described in Section~\ref{subsec:studied_data_aug} to generate new descriptions for the investigations. The original and augmented descriptions were used to train different SV assessment models (see Section~\ref{subsec:studied_sv_assessment_models}). These models were evaluated following the realistic setting of time-based evaluation rounds in Section~\ref{subsec:evaluation}.

\subsection{Research Questions}
\label{subsec:rqs}

We set out to answer the following two Research Questions (RQs) to distill the effectiveness of data augmentation for different SV assessment tasks using SV descriptions.

\begin{itemize}
[noitemsep,topsep=0pt,leftmargin=*]
  \item \textbf{RQ1: What is the significance of addressing data imbalance for the SV assessment tasks?} Existing studies have mostly developed SV assessment models without considering/mitigating the data imbalance issue illustrated in Section~\ref{subsec:data_aug_bg}. The existence of the data imbalance in the CVSS metrics is evident, but the impacts of the issue on respective SV assessment models are still largely unknown.
  For each of the seven CVSS metrics, RQ1 compares the performance of SV assessment models using descriptions generated by different data augmentation techniques with that of the baseline without using data augmentation.
  For each metric, if the performance of the model with data augmentation is better than that of the baseline, we can infer that data imbalance indeed negatively affects the task. Otherwise, the impact is of negligible concern.
  The findings of RQ1 are expected to demonstrate the importance of mitigating the data imbalance issue when developing CVSS-based SV assessment models.
  \item \textbf{RQ2: Which data augmentation techniques are effective for SV assessment?}
  While RQ1 shows the overall impact of data imbalance on the CVSS assessment metrics, RQ2 provides a more fine-grained analysis of the effectiveness of each data augmentation technique, i.e., whether a technique performs better or worse than the baseline on average.
  We also identify the optimal data augmentation technique for each SV assessment task.
  Moreover, RQ2 shows the extent to which the commonly used types of SV assessment models would benefit performance-wise from the data augmentation techniques for the tasks.
  The findings of RQ2 can inform the choice of particular data augmentation technique(s) to tackle data imbalance for SV assessment, which in turn opens up various future opportunities for improving the performance of the tasks in general.
\end{itemize}

\subsection{Dataset}
\label{subsec:dataset}

\begin{figure}[t]
  \centering
  \includegraphics[trim={19.1cm 0.35cm 22.2cm 0.5cm},clip,width=0.98\columnwidth,keepaspectratio]{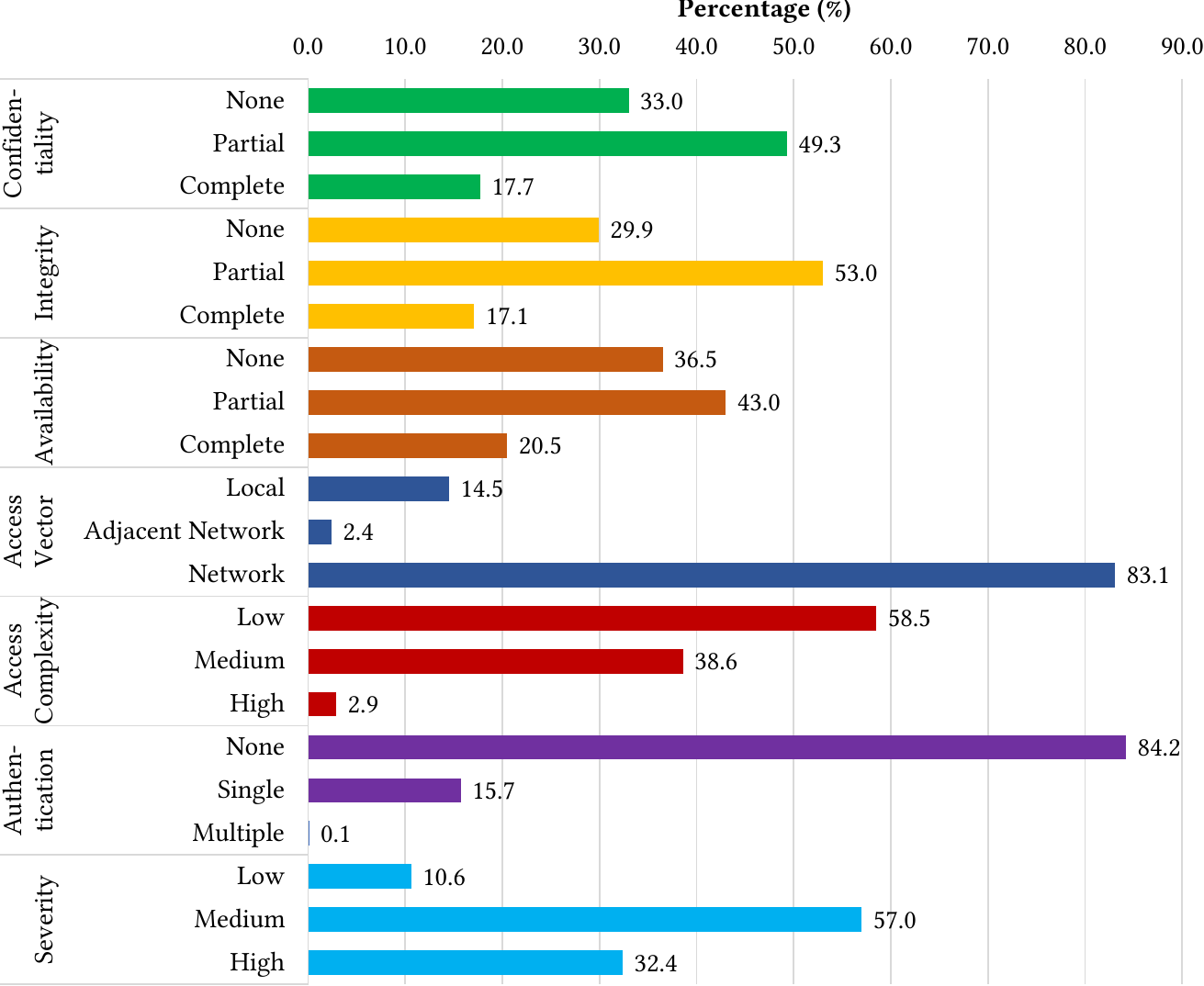}
  \caption{Data class distributions of the seven CVSS metrics used for SV assessment. \textbf{Note}: The total number of the collected SVs is 180,087.}
  \label{fig:cvss2_classes}
\end{figure}

To build a dataset for SV assessment, we collected real-world SVs reported on NVD~\cite{nvd_website} from 1988 to 2023.
To increase the relevance and reliability of our experiments, we discarded SVs that contained ``** REJECT **'' in their descriptions because they had been confirmed duplicated or incorrect by security experts.
We also excluded SVs that did not have any CVSS metrics.
Finally, we obtained a dataset consisting of \textit{180,087} SVs along with the respective expert-vetted descriptions and the values of the seven CVSS assessment metrics, i.e., Access Vector, Access Complexity, Authentication, Confidentiality, Integrity, Availability, and Severity.
The data class distributions of the collected CVSS metrics are given in \fig~\ref{fig:cvss2_classes}.
The values from this figure reinforce the earlier argument in Section~\ref{subsec:data_aug_bg} that data used for SV assessment based on the CVSS metrics are highly imbalanced and may negatively affect the performance of downstream models.

\begin{table*}[t]
\fontsize{7}{8}\selectfont
\caption{Examples of the original description and the descriptions generated by the studied data augmentation techniques of the CVE-1999-0315 SV. \textbf{Note}: The augmented descriptions were manually selected among multiple runs to enhance readability.}
\label{tab:da_examples}
\begin{tabular}{p{2.2cm}|p{2.2cm}|p{5.4cm}|p{6.5cm}}
\hline
\textbf{Category} & \textbf{Data Augmentation} & \textbf{Software Vulnerability Description} & \textbf{Explanation of the Changes} \\
\hline
\multicolumn{2}{c|}{None} & Buffer overflow in Solaris fdformat command gives root access to local users. & Original SV description \\
\hline
 \multirowcell{5}[0ex][l]{\textbf{Simple}\\ \textbf{Text Augmentation}}
& Insertion & Buffer overflow in Solaris fdformat command gives vulnerability root access to local users. & Inserting the word ``vulnerability''; vulnerability is a common word appearing in different SV descriptions. \\
\cline{2-4}
& Deletion & overflow in Solaris fdformat command gives root access to local users. & Removing the word ``Buffer'' \\
\cline{2-4}
& Substitution & Buffer error in Solaris fdformat command gives root access to local users. & Substituting/replacing the word ``overflow'' with the word ``error'' \\
\cline{2-4}
& Synonym\newline Replacement & Buffer overflow in Solaris fdformat command gives root path to local users. & Replacing the word ``access'' with the synonymous word ``path'' \\
\cline{2-4}
& Combination & error in Solaris fdformat command gives vulnerability root access to local users. & Inserting the word ``vulnerability'' + Substituting the word ``overflow'' with the word ``error'' + Removing the word ``Buffer'' \\
\hline
\hline
\multirowcell{5}[0ex][l]{\textbf{Contextual}\\ \textbf{Text Augmentation}} & Back Translation & Buffer overflow in Solaris fdformat command gives local users the root access. & Translating the original description to German, i.e., ``\textit{pufferüberlauf im fehler solari fdformat gibt lokal benutzern rootzugriff}'', and then translating the German description back to English \\
\cline{2-4}
& Paraphrasing & Solaris fdformat command allows users to access root via the buffer overflow bug, posing imminent threats. & Rewriting the original description while retaining the key semantics of the description \\
\hline
\end{tabular}
\end{table*}

\subsection{Studied Data Augmentation Techniques}
\label{subsec:studied_data_aug}

We studied three types of data augmentation techniques that work directly on the textual SV descriptions extracted in Section~\ref{subsec:dataset}: (\textit{i}) Data Sampling, (\textit{ii}) Simple Text Augmentation, and (\textit{iii}) Contextual Text Augmentation.
These techniques generated new/augmented descriptions that share the same labels as the respective original descriptions to balance the number of samples of all the classes of SV assessment tasks, i.e., CVSS metrics, aiming to address data imbalance shown in Section~\ref{subsec:sv_assessment_bg}.
We focused on the data augmentation techniques whose output is real text; we did not consider techniques that operate on the feature level like SMOTE~\cite{chawla2002smote} or on the model level like mixup~\cite{zhang2017mixup} because it is non-trivial to reconstruct real text from their output, hindering their interpretability.

\subsubsection{\textbf{Data Sampling}} 
\textit{Data Sampling} is the first type of data augmentation we employed for balancing the class distributions of CVSS metrics.
Specifically, we investigated two data sampling strategies: Random Over-sampling and Random Under-Sampling.
\textit{\textbf{Random Over-Sampling}} added random duplicates of the existing samples from minority classes so that the numbers of all the classes were equal for each of the SV assessment metrics.
Conversely, \textit{\textbf{Random Under-Sampling}} randomly removed the existing samples of the majority classes to match the number of the smallest class of each metric, i.e., the one with the least number of samples.
It is important to note that Random Under-Sampling does not directly align with the definition of data augmentation (i.e., adding new data), yet we still included it for the sake of completeness as it is also data sampling and has been used for SV assessment~\cite{han2017learning}.

\subsubsection{\textbf{Simple Text Augmentation}}
Unlike \textit{Data Sampling}, which only duplicated existing samples without making textual changes, \textit{Simple Text Augmentation} created new SV descriptions by modifying original SV descriptions (see \tab~\ref{tab:da_examples}). We considered the following textual modifications: (\textit{i}) Insertion, (\textit{ii}) Deletion, (\textit{iii}) Substitution, (\textit{iv}) Synonym Replacement, and (\textit{v}) Combination.
These operations have been commonly used to augment text in the Natural Language Processing domain (e.g.,~\cite{wei2019eda,bayer2022survey}), which is directly relevant to augmenting SV descriptions investigated in this study.
\textit{\textbf{Insertion}} created new descriptions by adding new word(s) at random position(s) in each description. We selected the frequent words, i.e., appearing in at least 0.1\% of descriptions in training set and not in the list of stop words, to add to the generated samples.
The threshold ensured the impact of the inserted words on model training was non-negligible~\cite{spanos2018multi}, without limiting the diversity of augmentations.
\textit{\textbf{Deletion}} randomly removed word(s) from each description to generate the respective augmented version.
\textit{\textbf{Substitution}} combined Insertion and Deletion, in which random word(s) in each description were first removed and then replaced with other frequent word(s) in the vocabulary of training set (excluding stop words), similar to Insertion.
\textit{\textbf{Synonym Replacement}} replaced word(s) in each description that had at least one synonymous alternative in WordNet~\cite{miller1995wordnet}; e.g., ``\textit{access}'' is replaced with ``\textit{path}'' in \tab~\ref{tab:da_examples}.
We prioritized the synonyms that frequently appeared in training set (excluding stop words) to ensure these words were properly captured during model training.
Lastly, \textit{\textbf{Combination}} performed a random combination of all of the above operations altogether to make changes to SV descriptions.
The techniques in \textit{Simple Text Augmentation} were implemented using the \code{nlpaug} library~\cite{ma2019nlpaug}.
It is worth noting that we applied these operations randomly to SV descriptions to increase the diversity of generated samples, yet we only changed up to 20\% of the words per description to avoid significant changes to the original semantics, as per the existing practice~\cite{wei2019eda}. We found changing 20+\% of the words tended to decrease model performance.
We would ensure to apply the operations to one word in an SV description in case it contained fewer than five words.
We did not swap the order of the words as this operation did not affect the \code{TF-IDF} feature extractor in Section~\ref{subsec:studied_sv_assessment_models}.

\subsubsection{\textbf{Contextual Text Augmentation}}
\textit{Simple Text Augmentation} treated each word independently without the surrounding context/words, which might not be optimal for preserving the semantics of an entire description.
\textit{Contextual Text Augmentation}, on the other hand, aimed at modifying the syntactic structure yet (theoretically) retaining the overall meaning of a description. We studied two representative techniques incorporating the context of SV descriptions: Back Translation and Paraphrasing.
These two techniques have been commonly used for text augmentation (e.g.,~\cite{sennrich2016improving,dai2023chataug,bayer2022survey}).
\textit{\textbf{Back Translation}} first translated/converted text to an intermediate language and then translated that back to the original language (English). The changes/augmentations in text mainly came from the variations in the two steps of translation.
We chose German as the intermediate language and the respective models for translating between German and English because they have been shown to be effective for these translation tasks~\cite{sennrich2016improving}. We also tested with another popular language, i.e., French, but the performance was worse.
\textit{\textbf{Paraphrasing}} rewrote a sentence using different words and/or text structures, while maintaining the original meaning. Our study used GPT~\cite{brown2020language}, the recent advance in Generative AI and used for SV tasks~\cite{fu2023chatgpt,le2024software}, to paraphrase SV descriptions with the prompt ``\textit{As a software security expert, please paraphrase the following text: \code{text}}''.
We also tried other prompts based on the recommended practices in the literature~\cite{dai2023chataug}, but the paraphrasing outputs were largely the same.
We implemented the aforementioned GPT-based paraphrasing using the \code{GPT4All} library~\cite{anand2023gpt4all}.

\subsection{Studied SV Assessment Models}
\label{subsec:studied_sv_assessment_models}

We leveraged the original and augmented SV descriptions to develop SV assessment models. These models were based on the four most widely used model types for the task~\cite{le2022survey}.

\subsubsection{\textbf{Random Forest (RF) + TF-IDF model}} This model employed RF~\cite{ho1995random} to classify CVSS metrics.
The RF model used TF-IDF features, i.e., the multiplication between the term frequency (times a word appears in a document) and the inverse document frequency (times a word appears in all documents) of each word in the vocabulary of training set.
RF and TF-IDF have been frequently used as the baselines for SV assessment (e.g.,~\cite{spanos2017assessment,spanos2018multi,le2019automated}).
Similar to these prior studies, we tuned the RF classifier using the hyperparameters: \textit{no. of classifiers}: \{100, 300, 500\}, \textit{max depth}: \{3, 5, 7, 9\} and \textit{leaf nodes}: \{100, 200, 300\}.
For TF-IDF, we also used a vocabulary with words appearing in at least 0.1\% of all the documents used for training.
In addition, we preprocessed text before extracting features with TF-IDF, including removing stop words and punctuations, converting text to lowercase, and applying stemming (i.e., changing words to their root form)~\cite{porter1980algorithm}. We used the stop words from the \emph{scikit-learn}~\cite{pedregosa2011scikit} and \emph{nltk}~\cite{loper2002nltk} libraries.
Regarding the punctuations, we only removed the ones at the end of a sentence or the ones followed by space(s) to retain important/relevant words in the software/security domains such as ``\emph{file.c}'' or ``\emph{cross-site} (scripting)''. 

\subsubsection{\textbf{RF + Doc2Vec model}} This baseline used the RF classifier with the same configurations as \code{RF + TF-IDF}, but with a different feature extractor, namely Doc2Vec~\cite{le2014distributed}. Doc2Vec derived the representation of an SV description using the information from the constituent words.
Similar to word2vec used in SV assessment (e.g.,~\cite{han2017learning,gong2019joint}), Doc2Vec captured the surrounding contexts of words missing in TF-IDF.
Moreover, Doc2Vec improved upon word2vec by assigning different weights to words in a document rather than simply averaging the word-wise vectors adopted by word2vec.
Doc2Vec has also been previously used for SV assessment~\cite{kanakogi2021tracing}.
We followed the suggestion of Doc2Vec's original authors~\cite{le2014distributed} to train the Doc2Vec feature extraction using the distributed memory architecture.
For Doc2Vec, we also applied the same text-preprocessing steps used for TF-IDF.
As per the prior studies~\cite{le2019automated,han2017learning} and our preliminary analysis, we used the default configurations of Doc2Vec because other values did not change results significantly.

\subsubsection{\textbf{Convolutional Neural Network (CNN) model}}
CNN~\cite{kim2014convolutional} has been used widely for SV assessment (e.g.,~\cite{han2017learning,gong2019joint,sharma2021software,anwar2021cleaning}). The model started with an embedding layer connected to one convolutional layer.
We considered various embedding sizes of 100, 200, 300.
The convolutional layer had $F$ filters with pre-defined size $K$.
We tried different numbers of filters of 64, 128, 256, and different filter sizes of 1, 3, 5.
The hyperparameters of CNN were inspired by existing studies~\cite{han2017learning}.
These filters extracted patterns of phrases (consecutive words) of size $K$.
The outputs of the filters, went through $\operatorname{ReLU}(x) = \operatorname{max}(0, x)$, a non-linear activation function~\mbox{\cite{nair2010rectified}}.
We iterated through the convolutional process, moving filters sequentially down along the input vector from the beginning to the end, employing a stride of one. This smaller stride was chosen to capture highly detailed information within input descriptions, distinguishing it from larger strides.
The convolutional outputs entered a max-pooling layer to produce a vector representing an input description. The output of the pooling layer was then fed into a softmax layer to classify each CVSS metric.
We followed the existing practices~\cite{han2017learning} to train CNN for SV assessment.

\subsubsection{\textbf{Long-Short Term Memory (LSTM) model}}
LSTM~\cite{hochreiter1997long} has also been commonly used as an alternative to CNN to better capture the long-term dependencies among the words in SV descriptions~\cite{gong2019joint}. The LSTM model had the same embedding layer as the CNN model connected to a forward LSTM-based network that read the input from left to right.
We investigated different numbers of LSTM cells of 64, 128, and 256, similar to related studies (e.g.,~\cite{sahin2019conceptual,gong2019joint}).
After processing each description, the LSTM cells output a hidden vector representing the whole description. We fed such vector into a softmax layer to perform the classification of CVSS metrics, similar to CNN.
Note that we tried Bi-directional LSTM for SV assessment, but it did not yield a stronger performance.

\subsection{Evaluation Techniques and Metrics}
\label{subsec:evaluation}

\subsubsection{\textbf{Data splitting}}
We employed time-based data splits for training, validating, and testing our models. This setup simulated real-world scenarios where future unseen data was not included during training, as recommended in the literature~\cite{spanos2018multi,le2019automated}.
This approach involved three rounds of training, validation, and testing, utilizing five equally sized folds based on the published dates of SVs.
Each round $i$ utilized folds from 1 to $i$, $i+1$, and $i+2$ for training, validation, and testing, respectively. We applied data augmentation \textit{only} to the descriptions in the training set and used only the original descriptions for validation/testing.
We then selected the optimal model as the one with the highest average validation performance, based on the hyperparameters in Section~\ref{subsec:studied_sv_assessment_models}. The average testing performance of the optimal model was then reported, ensuring more stable outcomes than a single testing set~\cite{raschka2018model}.

\subsubsection{\textbf{Evaluation measures}}

To assess the performance of automated SV assessment, we applied the F1-Score and Matthews Correlation Coefficient (MCC) metrics, which have been widely employed for CVSS classification (e.g., \cite{han2017learning,spanos2018multi,jimenez2019importance,le2024latent}). These metrics were suitable for handling imbalanced classes within our data~\cite{luque2019impact}, as illustrated in \fig~\ref{fig:data_imbalance}. F1-Score spans between 0 to 1, while MCC ranges from –1 to 1, where 1 signifies the optimal value for both metrics. MCC, considering all the cells explicitly in a confusion matrix, was utilized to select the optimal models~\cite{luque2019impact}. Given that the tasks had multiple classes, we used macro F1-Score and the multi-class version of MCC~\cite{spanos2018multi,gorodkin2004comparing}. It is important to note that MCC does not have a direct correlation with F1-Score.

\noindent
\subsubsection{\textbf{Statistical analysis}}
To affirm the significance of our findings, we used the non-parametric Wilcoxon signed-rank test and its effect size ($r = Z / \sqrt{N}$, where $Z$ is the statistic score of the test and $N$ represents the total sample size). The magnitude of the effect size ($r \leq 0.1$: negligible, $0.1 < r \leq 0.3$: small, $0.3 < r \leq 0.5$: medium, $r > 0.5$: large) followed the established guidelines~\cite{field2013discovering}. We would confirm a test result statistically significant when the confidence level was over 99\%, equivalent to $p$-value $<$ 0.01.
We used this effect size because it has been commonly used for assessing and comparing defect/SV prediction results (e.g.,~\cite{wattanakriengkrai2020predicting,le2022use}).

\section{Experimental Results of Data Augmentation for SV Assessment}
\label{sec:results}

\begin{table*}[t]
\fontsize{7}{8}\selectfont
\caption{Testing performance in terms of MCC and F1-Score of the baseline None models (without using data augmentation) and the models using nine different data augmentation techniques. \textbf{Notes}: The baseline performance is highlighted in yellow. The best performance of the models using data augmentation is highlighted in dark green.}
\label{tab:da_results}

\begin{tabular}{c|l|cc|cc|cc|cc|cc|cc|cc}
 \hline
 &
  \multicolumn{1}{c|}{} &
  \multicolumn{2}{c|}{\textbf{Access Vector}} &
  \multicolumn{2}{c|}{\textbf{Access Comp.}} &
  \multicolumn{2}{c|}{\textbf{Authentication}} &
  \multicolumn{2}{c|}{\textbf{Confidentiality}} &
  \multicolumn{2}{c|}{\textbf{Integrity}} &
  \multicolumn{2}{c|}{\textbf{Availability}} &
  \multicolumn{2}{c}{\textbf{Severity}} \\
  \cline{3-16}
\multirow{-2}{*}{\textbf{Model}} &
  \multicolumn{1}{c|}{\multirow{-2}{*}{\textbf{Data Augmentation}}} &
  \textbf{MCC} &
  \textbf{F1} &
  \textbf{MCC} &
  \textbf{F1} &
  \textbf{MCC} &
  \textbf{F1} &
  \textbf{MCC} &
  \textbf{F1} &
  \textbf{MCC} &
  \textbf{F1} &
  \textbf{MCC} &
  \textbf{F1} &
  \textbf{MCC} &
  \textbf{F1} \\ \hline
 &
  None &
  \cellcolor[HTML]{FFE699}0.458 &
  \cellcolor[HTML]{FFE699}0.532 &
  \cellcolor[HTML]{FFE699}0.604 &
  \cellcolor[HTML]{FFE699}0.530 &
  \cellcolor[HTML]{FFE699}0.311 &
  \cellcolor[HTML]{FFE699}0.378 &
  \cellcolor[HTML]{FFE699}0.509 &
  \cellcolor[HTML]{FFE699}0.638 &
  \cellcolor[HTML]{FFE699}0.531 &
  \cellcolor[HTML]{FFE699}0.656 &
  \cellcolor[HTML]{FFE699}0.507 &
  \cellcolor[HTML]{FFE699}0.660 &
  \cellcolor[HTML]{FFE699}0.327 &
  \cellcolor[HTML]{FFE699}0.470 \\
 &
  Over-Sampling &
  \cellcolor[HTML]{A9D08E}0.553 &
  \cellcolor[HTML]{A9D08E}0.632 &
  \cellcolor[HTML]{548235}\textcolor{white}{0.625} &
  \cellcolor[HTML]{548235}\textcolor{white}{0.609} &
  \cellcolor[HTML]{548235}\textcolor{white}{0.428} &
  \cellcolor[HTML]{548235}\textcolor{white}{0.537} &
  \cellcolor[HTML]{548235}\textcolor{white}{0.533} &
  \cellcolor[HTML]{548235}\textcolor{white}{0.667} &
  \cellcolor[HTML]{548235}\textcolor{white}{0.545} &
  \cellcolor[HTML]{548235}\textcolor{white}{0.658} &
  \cellcolor[HTML]{A9D08E}0.528 &
  \cellcolor[HTML]{A9D08E}0.667 &
  \cellcolor[HTML]{548235}\textcolor{white}{0.362} &
  \cellcolor[HTML]{548235}\textcolor{white}{0.565} \\
 &
  Under-Sampling &
  \cellcolor[HTML]{A9D08E}0.543 &
  \cellcolor[HTML]{A9D08E}0.614 &
  0.602 &
  \cellcolor[HTML]{A9D08E}0.589 &
  0.115 &
  0.368 &
  \cellcolor[HTML]{A9D08E}0.528 &
  \cellcolor[HTML]{A9D08E}0.664 &
  \cellcolor[HTML]{A9D08E}0.540 &
  0.653 &
  \cellcolor[HTML]{A9D08E}0.524 &
  \cellcolor[HTML]{A9D08E}0.665 &
  \cellcolor[HTML]{A9D08E}0.355 &
  \cellcolor[HTML]{A9D08E}0.555 \\
 &
  Insertion &
  \cellcolor[HTML]{A9D08E}0.540 &
  \cellcolor[HTML]{A9D08E}0.611 &
  \cellcolor[HTML]{A9D08E}0.625 &
  \cellcolor[HTML]{A9D08E}0.608 &
  \cellcolor[HTML]{A9D08E}0.420 &
  \cellcolor[HTML]{A9D08E}0.465 &
  \cellcolor[HTML]{A9D08E}0.532 &
  \cellcolor[HTML]{A9D08E}0.666 &
  \cellcolor[HTML]{A9D08E}0.540 &
  0.653 &
  0.472 &
  0.644 &
  \cellcolor[HTML]{A9D08E}0.357 &
  \cellcolor[HTML]{A9D08E}0.557 \\
 &
 Deletion &
  0.451 &
  \cellcolor[HTML]{A9D08E}0.536 &
  0.481 &
  0.523 &
  0.278 &
  \cellcolor[HTML]{A9D08E}0.380 &
  0.497 &
  0.619 &
  0.508 &
  0.614 &
  0.479 &
  0.646 &
  0.311 &
  0.461 \\
 &
  Substitution &
  0.441 &
  0.520 &
  0.460 &
  0.511 &
  0.253 &
  0.368 &
  0.488 &
  0.609 &
  0.506 &
  0.613 &
  0.478 &
  0.646 &
  0.306 &
  0.460 \\
 &
 Synonym Replacement &
  \cellcolor[HTML]{A9D08E}0.562 &
  \cellcolor[HTML]{A9D08E}0.634 &
  \cellcolor[HTML]{A9D08E}0.608 &
  \cellcolor[HTML]{A9D08E}0.584 &
  \cellcolor[HTML]{A9D08E}0.398 &
  \cellcolor[HTML]{A9D08E}0.465 &
  \cellcolor[HTML]{A9D08E}0.526 &
  \cellcolor[HTML]{A9D08E}0.657 &
  \cellcolor[HTML]{A9D08E}0.535 &
  0.648 &
  \cellcolor[HTML]{A9D08E}0.525 &
  \cellcolor[HTML]{A9D08E}0.666 &
  \cellcolor[HTML]{A9D08E}0.344 &
  \cellcolor[HTML]{A9D08E}0.541 \\
 &
  Combination &
  \cellcolor[HTML]{548235}\textcolor{white}{0.573} &
  \cellcolor[HTML]{548235}\textcolor{white}{0.647} &
  \cellcolor[HTML]{A9D08E}0.620 &
  \cellcolor[HTML]{A9D08E}0.596 &
  \cellcolor[HTML]{A9D08E}0.406 &
  \cellcolor[HTML]{A9D08E}0.474 &
  \cellcolor[HTML]{A9D08E}0.530 &
  \cellcolor[HTML]{A9D08E}0.660 &
  \cellcolor[HTML]{A9D08E}0.539 &
  \cellcolor[HTML]{A9D08E}0.650 &
  \cellcolor[HTML]{548235}\textcolor{white}{0.535} &
  \cellcolor[HTML]{548235}\textcolor{white}{0.679} &
  \cellcolor[HTML]{A9D08E}0.351 &
  \cellcolor[HTML]{A9D08E}0.552 \\
 &
  Back Translation &
  0.149 &
  0.190 &
  0.140 &
  0.178 &
  0.062 &
  0.153 &
  0.157 &
  0.191 &
  0.192 &
  0.238 &
  0.187 &
  0.215 &
  0.089 &
  0.162 \\
\multirow{-10}{*}{\rotatebox[origin=c]{90}{RF + TF-IDF}} &
  Paraphrasing &
  0.396 &
  0.481 &
  0.403 &
  0.476 &
  0.181 &
  \cellcolor[HTML]{A9D08E}0.465 &
  0.461 &
  0.588 &
  0.495 &
  0.612 &
  0.480 &
  0.646 &
  0.297 &
  \cellcolor[HTML]{A9D08E}0.477 \\ \hline
 &
  None &
  \cellcolor[HTML]{FFE699}0.134 &
  \cellcolor[HTML]{FFE699}0.302 &
  \cellcolor[HTML]{FFE699}0.154 &
  \cellcolor[HTML]{FFE699}0.307 &
  \cellcolor[HTML]{FFE699}0.148 &
  \cellcolor[HTML]{FFE699}0.327 &
  \cellcolor[HTML]{FFE699}0.231 &
  \cellcolor[HTML]{FFE699}0.354 &
  \cellcolor[HTML]{FFE699}0.212 &
  \cellcolor[HTML]{FFE699}0.338 &
  \cellcolor[HTML]{FFE699}0.202 &
  \cellcolor[HTML]{FFE699}0.368 &
  \cellcolor[HTML]{FFE699}0.147 &
  \cellcolor[HTML]{FFE699}0.373 \\
 &
  Over-Sampling &
  \cellcolor[HTML]{548235}\textcolor{white}{0.243} &
  \cellcolor[HTML]{548235}\textcolor{white}{0.424} &
  \cellcolor[HTML]{548235}\textcolor{white}{0.200} &
  \cellcolor[HTML]{548235}\textcolor{white}{0.379} &
  \cellcolor[HTML]{548235}\textcolor{white}{0.211} &
  \cellcolor[HTML]{548235}\textcolor{white}{0.465} &
  \cellcolor[HTML]{548235}\textcolor{white}{0.256} &
  \cellcolor[HTML]{548235}\textcolor{white}{0.462} &
  \cellcolor[HTML]{548235}\textcolor{white}{0.247} &
  \cellcolor[HTML]{A9D08E}0.412 &
  \cellcolor[HTML]{A9D08E}0.212 &
  \cellcolor[HTML]{A9D08E}0.438 &
  \cellcolor[HTML]{A9D08E}0.168 &
  \cellcolor[HTML]{548235}\textcolor{white}{0.425} \\
 &
  Under-Sampling &
  \cellcolor[HTML]{A9D08E}0.140 &
  0.293 &
  0.140 &
  0.290 &
  0.142 &
  0.313 &
  0.224 &
  \cellcolor[HTML]{A9D08E}0.430 &
  \cellcolor[HTML]{A9D08E}0.230 &
  \cellcolor[HTML]{A9D08E}0.410 &
  \cellcolor[HTML]{A9D08E}0.218 &
  \cellcolor[HTML]{A9D08E}0.393 &
  0.130 &
  0.322 \\
 &
  Insertion &
  0.114 &
  \cellcolor[HTML]{A9D08E}0.374 &
  \cellcolor[HTML]{A9D08E}0.155 &
  0.296 &
  0.138 &
  0.305 &
  0.226 &
  \cellcolor[HTML]{A9D08E}0.446 &
  0.195 &
  \cellcolor[HTML]{A9D08E}0.452 &
  0.193 &
  \cellcolor[HTML]{A9D08E}0.440 &
  \cellcolor[HTML]{A9D08E}0.174 &
  \cellcolor[HTML]{A9D08E}0.400 \\
 &
 Deletion &
  \cellcolor[HTML]{A9D08E}0.167 &
  \cellcolor[HTML]{A9D08E}0.387 &
  0.145 &
  0.296 &
  \cellcolor[HTML]{A9D08E}0.149 &
  \cellcolor[HTML]{A9D08E}0.328 &
  0.225 &
  \cellcolor[HTML]{A9D08E}0.428 &
  0.206 &
  \cellcolor[HTML]{A9D08E}0.445 &
  0.180 &
  \cellcolor[HTML]{A9D08E}0.429 &
  \cellcolor[HTML]{A9D08E}0.174 &
  \cellcolor[HTML]{A9D08E}0.400 \\
 &
  Substitution &
  \cellcolor[HTML]{A9D08E}0.161 &
  \cellcolor[HTML]{A9D08E}0.398 &
  \cellcolor[HTML]{A9D08E}0.157 &
  0.302 &
  0.139 &
  0.306 &
  0.209 &
  \cellcolor[HTML]{A9D08E}0.453 &
  0.186 &
  \cellcolor[HTML]{A9D08E}0.451 &
  0.197 &
  \cellcolor[HTML]{A9D08E}0.421 &
  \cellcolor[HTML]{A9D08E}0.183 &
  \cellcolor[HTML]{A9D08E}0.382 \\
 &
 Synonym Replacement &
  \cellcolor[HTML]{A9D08E}0.215 &
  \cellcolor[HTML]{A9D08E}0.408 &
  \cellcolor[HTML]{A9D08E}0.196 &
  \cellcolor[HTML]{A9D08E}0.317 &
  0.139 &
  0.307 &
  0.221 &
  \cellcolor[HTML]{A9D08E}0.419 &
  \cellcolor[HTML]{A9D08E}0.215 &
  \cellcolor[HTML]{A9D08E}0.452 &
  \cellcolor[HTML]{A9D08E}0.208 &
  \cellcolor[HTML]{A9D08E}0.413 &
  \cellcolor[HTML]{A9D08E}0.170 &
  \cellcolor[HTML]{A9D08E}0.394 \\
 &
  Combination &
  \cellcolor[HTML]{A9D08E}0.219 &
  \cellcolor[HTML]{A9D08E}0.417 &
  \cellcolor[HTML]{A9D08E}0.198 &
  \cellcolor[HTML]{A9D08E}0.323 &
  \cellcolor[HTML]{A9D08E}0.167 &
  \cellcolor[HTML]{A9D08E}0.368 &
  \cellcolor[HTML]{A9D08E}0.240 &
  \cellcolor[HTML]{A9D08E}0.441 &
  \cellcolor[HTML]{A9D08E}0.220 &
  \cellcolor[HTML]{548235}\textcolor{white}{0.461} &
  \cellcolor[HTML]{548235}\textcolor{white}{0.221} &
  \cellcolor[HTML]{548235}\textcolor{white}{0.443} &
  \cellcolor[HTML]{548235}\textcolor{white}{0.189} &
  \cellcolor[HTML]{A9D08E}0.401 \\
 &
  Back Translation &
  0.037 &
  0.064 &
  0.035 &
  0.090 &
  0.030 &
  0.052 &
  0.039 &
  0.076 &
  0.051 &
  0.059 &
  0.048 &
  0.098 &
  0.021 &
  0.063 \\
\multirow{-10}{*}{\rotatebox[origin=c]{90}{RF + Doc2Vec}} &
  Paraphrasing &
  \cellcolor[HTML]{A9D08E}0.196 &
  \cellcolor[HTML]{A9D08E}0.313 &
  0.148 &
  \cellcolor[HTML]{A9D08E}0.376 &
  0.148 &
  0.326 &
  0.195 &
  \cellcolor[HTML]{A9D08E}0.384 &
  0.208 &
  \cellcolor[HTML]{A9D08E}0.381 &
  \cellcolor[HTML]{A9D08E}0.212 &
  \cellcolor[HTML]{A9D08E}0.419 &
  0.095 &
  0.358 \\ \hline
 &
  None &
  \cellcolor[HTML]{FFE699}0.578 &
  \cellcolor[HTML]{FFE699}0.573 &
  \cellcolor[HTML]{FFE699}0.613 &
  \cellcolor[HTML]{FFE699}0.589 &
  \cellcolor[HTML]{FFE699}0.465 &
  \cellcolor[HTML]{FFE699}0.691 &
  \cellcolor[HTML]{FFE699}0.574 &
  \cellcolor[HTML]{FFE699}0.685 &
  \cellcolor[HTML]{FFE699}0.592 &
  \cellcolor[HTML]{FFE699}0.691 &
  \cellcolor[HTML]{FFE699}0.540 &
  \cellcolor[HTML]{FFE699}0.660 &
  \cellcolor[HTML]{FFE699}0.338 &
  \cellcolor[HTML]{FFE699}0.535 \\
 &
  Over-Sampling &
  \cellcolor[HTML]{A9D08E}0.584 &
  0.570 &
  0.611 &
  0.544 &
  0.410 &
  0.453 &
  0.556 &
  0.677 &
  0.561 &
  0.688 &
  \cellcolor[HTML]{A9D08E}0.552 &
  \cellcolor[HTML]{A9D08E}0.673 &
  0.336 &
  \cellcolor[HTML]{A9D08E}0.559 \\
 &
  Under-Sampling &
  0.491 &
  \cellcolor[HTML]{A9D08E}0.583 &
  0.445 &
  0.508 &
  0.183 &
  0.384 &
  0.560 &
  0.681 &
  0.581 &
  0.688 &
  \cellcolor[HTML]{A9D08E}0.547 &
  \cellcolor[HTML]{A9D08E}0.670 &
  \cellcolor[HTML]{A9D08E}0.358 &
  \cellcolor[HTML]{A9D08E}0.572 \\
 &
  Insertion &
  \cellcolor[HTML]{A9D08E}0.620 &
  \cellcolor[HTML]{A9D08E}0.597 &
  \cellcolor[HTML]{A9D08E}0.630 &
  0.564 &
  0.437 &
  0.678 &
  \cellcolor[HTML]{A9D08E}0.580 &
  0.681 &
  \cellcolor[HTML]{A9D08E}0.610 &
  \cellcolor[HTML]{A9D08E}0.708 &
  \cellcolor[HTML]{A9D08E}0.572 &
  \cellcolor[HTML]{A9D08E}0.681 &
  \cellcolor[HTML]{A9D08E}0.352 &
  \cellcolor[HTML]{A9D08E}0.556 \\
 &
 Deletion &
  \cellcolor[HTML]{A9D08E}0.594 &
  \cellcolor[HTML]{A9D08E}0.593 &
  \cellcolor[HTML]{A9D08E}0.641 &
  0.564 &
  \cellcolor[HTML]{A9D08E}0.474 &
  \cellcolor[HTML]{A9D08E}0.701 &
  \cellcolor[HTML]{A9D08E}0.584 &
  0.676 &
  \cellcolor[HTML]{A9D08E}0.607 &
  \cellcolor[HTML]{A9D08E}0.702 &
  \cellcolor[HTML]{A9D08E}0.564 &
  \cellcolor[HTML]{A9D08E}0.683 &
  \cellcolor[HTML]{A9D08E}0.358 &
  \cellcolor[HTML]{A9D08E}0.570 \\
 &
  Substitution &
  \cellcolor[HTML]{A9D08E}0.627 &
  \cellcolor[HTML]{A9D08E}0.620 &
  \cellcolor[HTML]{A9D08E}0.620 &
  0.554 &
  \cellcolor[HTML]{548235}\textcolor{white}{0.491} &
  \cellcolor[HTML]{548235}\textcolor{white}{0.732} &
  0.571 &
  \cellcolor[HTML]{A9D08E}0.690 &
  \cellcolor[HTML]{A9D08E}0.612 &
  \cellcolor[HTML]{A9D08E}0.699 &
  \cellcolor[HTML]{A9D08E}0.559 &
  \cellcolor[HTML]{A9D08E}0.677 &
  \cellcolor[HTML]{A9D08E}0.372 &
  \cellcolor[HTML]{A9D08E}0.572 \\
 &
 Synonym Replacement &
  \cellcolor[HTML]{A9D08E}0.619 &
  \cellcolor[HTML]{A9D08E}0.600 &
  \cellcolor[HTML]{A9D08E}0.633 &
  0.546 &
  \cellcolor[HTML]{A9D08E}0.489 &
  \cellcolor[HTML]{A9D08E}0.714 &
  \cellcolor[HTML]{A9D08E}0.582 &
  \cellcolor[HTML]{A9D08E}0.694 &
  \cellcolor[HTML]{A9D08E}0.601 &
  \cellcolor[HTML]{A9D08E}0.701 &
  \cellcolor[HTML]{A9D08E}0.569 &
  \cellcolor[HTML]{A9D08E}0.681 &
  0.334 &
  \cellcolor[HTML]{A9D08E}0.558 \\
 &
  Combination &
  \cellcolor[HTML]{548235}\textcolor{white}{0.633} &
  \cellcolor[HTML]{548235}\textcolor{white}{0.627} &
  \cellcolor[HTML]{548235}\textcolor{white}{0.646} &
  \cellcolor[HTML]{548235}\textcolor{white}{0.629} &
  \cellcolor[HTML]{A9D08E}0.479 &
  \cellcolor[HTML]{A9D08E}0.701 &
  \cellcolor[HTML]{548235}\textcolor{white}{0.586} &
  \cellcolor[HTML]{548235}\textcolor{white}{0.701} &
  \cellcolor[HTML]{548235}\textcolor{white}{0.613} &
  \cellcolor[HTML]{548235}\textcolor{white}{0.712} &
  \cellcolor[HTML]{548235}\textcolor{white}{0.574} &
  \cellcolor[HTML]{548235}\textcolor{white}{0.684} &
  \cellcolor[HTML]{548235}\textcolor{white}{0.374} &
  \cellcolor[HTML]{548235}\textcolor{white}{0.589} \\
 &
  Back Translation &
  0.327 &
  0.405 &
  0.362 &
  0.357 &
  0.270 &
  0.299 &
  0.329 &
  0.411 &
  0.388 &
  0.471 &
  0.335 &
  0.392 &
  0.180 &
  0.366 \\
\multirow{-10}{*}{\rotatebox[origin=c]{90}{CNN}} &
  Paraphrasing &
  0.536 &
  \cellcolor[HTML]{A9D08E}0.606 &
  0.570 &
  0.540 &
  0.400 &
  0.461 &
  0.528 &
  0.656 &
  0.576 &
  0.686 &
  0.513 &
  0.644 &
  0.299 &
  0.530 \\ \hline
 &
  None &
  \cellcolor[HTML]{FFE699}0.585 &
  \cellcolor[HTML]{FFE699}0.592 &
  \cellcolor[HTML]{FFE699}0.589 &
  \cellcolor[HTML]{FFE699}0.532 &
  \cellcolor[HTML]{FFE699}0.500 &
  \cellcolor[HTML]{FFE699}0.700 &
  \cellcolor[HTML]{FFE699}0.570 &
  \cellcolor[HTML]{FFE699}0.672 &
  \cellcolor[HTML]{FFE699}0.583 &
  \cellcolor[HTML]{FFE699}0.683 &
  \cellcolor[HTML]{FFE699}0.544 &
  \cellcolor[HTML]{FFE699}0.659 &
  \cellcolor[HTML]{FFE699}0.342 &
  \cellcolor[HTML]{FFE699}0.539 \\
 &
  Over-Sampling &
  \cellcolor[HTML]{A9D08E}0.600 &
  0.579 &
  \cellcolor[HTML]{A9D08E}0.617 &
  \cellcolor[HTML]{A9D08E}0.544 &
  0.401 &
  0.446 &
  \cellcolor[HTML]{A9D08E}0.585 &
  \cellcolor[HTML]{A9D08E}0.699 &
  0.571 &
  \cellcolor[HTML]{A9D08E}0.687 &
  \cellcolor[HTML]{A9D08E}0.549 &
  \cellcolor[HTML]{A9D08E}0.671 &
  \cellcolor[HTML]{A9D08E}0.343 &
  \cellcolor[HTML]{A9D08E}0.565 \\
 &
  Under-Sampling &
  0.492 &
  0.574 &
  0.450 &
  0.514 &
  0.135 &
  0.359 &
  0.541 &
  0.671 &
  0.572 &
  \cellcolor[HTML]{A9D08E}0.687 &
  \cellcolor[HTML]{A9D08E}0.546 &
  \cellcolor[HTML]{A9D08E}0.668 &
  \cellcolor[HTML]{A9D08E}0.344 &
  \cellcolor[HTML]{A9D08E}0.561 \\
 &
  Insertion &
  \cellcolor[HTML]{A9D08E}0.647 &
  \cellcolor[HTML]{A9D08E}0.656 &
  \cellcolor[HTML]{A9D08E}0.648 &
  \cellcolor[HTML]{A9D08E}0.555 &
  \cellcolor[HTML]{A9D08E}0.508 &
  \cellcolor[HTML]{A9D08E}0.728 &
  \cellcolor[HTML]{A9D08E}0.590 &
  \cellcolor[HTML]{A9D08E}0.699 &
  \cellcolor[HTML]{A9D08E}0.596 &
  \cellcolor[HTML]{A9D08E}0.693 &
  \cellcolor[HTML]{A9D08E}0.555 &
  \cellcolor[HTML]{A9D08E}0.674 &
  \cellcolor[HTML]{A9D08E}0.364 &
  \cellcolor[HTML]{A9D08E}0.583 \\
 &
 Deletion &
  \cellcolor[HTML]{A9D08E}0.634 &
  \cellcolor[HTML]{A9D08E}0.632 &
  \cellcolor[HTML]{A9D08E}0.639 &
  \cellcolor[HTML]{A9D08E}0.552 &
  0.479 &
  \cellcolor[HTML]{A9D08E}0.708 &
  \cellcolor[HTML]{A9D08E}0.580 &
  \cellcolor[HTML]{A9D08E}0.686 &
  \cellcolor[HTML]{A9D08E}0.612 &
  \cellcolor[HTML]{A9D08E}0.700 &
  \cellcolor[HTML]{A9D08E}0.568 &
  \cellcolor[HTML]{A9D08E}0.681 &
  \cellcolor[HTML]{A9D08E}0.353 &
  \cellcolor[HTML]{A9D08E}0.581 \\
 &
  Substitution &
  \cellcolor[HTML]{A9D08E}0.638 &
  \cellcolor[HTML]{A9D08E}0.650 &
  \cellcolor[HTML]{A9D08E}0.644 &
  \cellcolor[HTML]{A9D08E}0.568 &
  0.465 &
  \cellcolor[HTML]{A9D08E}0.704 &
  \cellcolor[HTML]{A9D08E}0.586 &
  \cellcolor[HTML]{A9D08E}0.696 &
  \cellcolor[HTML]{A9D08E}0.609 &
  \cellcolor[HTML]{A9D08E}0.708 &
  \cellcolor[HTML]{A9D08E}0.558 &
  \cellcolor[HTML]{A9D08E}0.680 &
  \cellcolor[HTML]{A9D08E}0.345 &
  \cellcolor[HTML]{A9D08E}0.568 \\
 &
 Synonym Replacement &
  \cellcolor[HTML]{A9D08E}0.637 &
  \cellcolor[HTML]{A9D08E}0.634 &
  \cellcolor[HTML]{548235}\textcolor{white}{0.663} &
  \cellcolor[HTML]{548235}\textcolor{white}{0.647} &
  \cellcolor[HTML]{A9D08E}0.514 &
  \cellcolor[HTML]{A9D08E}0.741 &
  \cellcolor[HTML]{A9D08E}0.589 &
  \cellcolor[HTML]{A9D08E}0.689 &
  \cellcolor[HTML]{A9D08E}0.604 &
  \cellcolor[HTML]{A9D08E}0.697 &
  \cellcolor[HTML]{A9D08E}0.563 &
  \cellcolor[HTML]{A9D08E}0.685 &
  \cellcolor[HTML]{548235}\textcolor{white}{0.381} &
  \cellcolor[HTML]{548235}\textcolor{white}{0.600} \\
 &
  Combination &
  \cellcolor[HTML]{548235}\textcolor{white}{0.650} &
  \cellcolor[HTML]{548235}\textcolor{white}{0.675} &
  \cellcolor[HTML]{A9D08E}0.633 &
  \cellcolor[HTML]{A9D08E}0.605 &
  \cellcolor[HTML]{548235}\textcolor{white}{0.522} &
  \cellcolor[HTML]{548235}\textcolor{white}{0.743} &
  \cellcolor[HTML]{548235}\textcolor{white}{0.592} &
  \cellcolor[HTML]{548235}\textcolor{white}{0.700} &
  \cellcolor[HTML]{548235}\textcolor{white}{0.619} &
  \cellcolor[HTML]{548235}\textcolor{white}{0.711} &
  \cellcolor[HTML]{548235}\textcolor{white}{0.580} &
  \cellcolor[HTML]{548235}\textcolor{white}{0.686} &
  \cellcolor[HTML]{A9D08E}0.369 &
  \cellcolor[HTML]{A9D08E}0.596 \\
 &
  Back Translation &
  0.236 &
  0.261 &
  0.254 &
  0.244 &
  0.189 &
  0.229 &
  0.268 &
  0.294 &
  0.267 &
  0.295 &
  0.217 &
  0.299 &
  0.139 &
  0.259 \\
\multirow{-10}{*}{\rotatebox[origin=c]{90}{LSTM}} &
  Paraphrasing &
  0.573 &
  \cellcolor[HTML]{A9D08E}0.617 &
  \cellcolor[HTML]{A9D08E}0.591 &
  \cellcolor[HTML]{A9D08E}0.568 &
  0.416 &
  0.467 &
  0.537 &
  0.664 &
  0.576 &
  \cellcolor[HTML]{A9D08E}0.686 &
  0.541 &
  \cellcolor[HTML]{A9D08E}0.660 &
  0.311 &
  \cellcolor[HTML]{A9D08E}0.551 \\ \hline
  \multicolumn{2}{c|}{\textbf{Avg. \% of Best Improvements}} & 31.8 & 21.4 & 12.9 & 16.7 & 22.4 & 24.1 & 5.3 & 10.4 & 7.2 & 11.0 & 7.0 & 7.7 & 15.5 & 13.8 \\
  \multicolumn{2}{c|}{\textbf{$p$-value of Best Improvements}} & 1.3e-4 & 2.3e-7 & 8.5e-7 & 3.4e-7 & 1.1e-3 & 2.7e-5 & 6.2e-3 & 1.6e-3 & 3.3e-3 & 5.1e-3 & 4.3e-5 & 8.5e-6 & 4.2e-7 & 2.3e-7 \\
  \multicolumn{2}{c|}{\textbf{Effect size of Best Improvements}} & 0.501 & 0.808 & 0.244 & 0.711 & 0.433 & 0.630 & 0.148 & 0.373 & 0.143 & 0.358 & 0.221 & 0.292 & 0.431 & 0.870 \\
  \hline
\end{tabular}%
\end{table*}

\subsection{RQ1: Significance of Mitigating Data Imbalance for SV Assessment Tasks}
\label{subsec:rq1_results}

As shown in \tab~\ref{tab:da_results}, addressing the data imbalance issue using data augmentation techniques could substantially improve the predictive performance of all the seven CVSS-based assessment tasks.
The best models using data augmentation, averaging all four model types in Section~\ref{subsec:studied_sv_assessment_models}, produced 5.3--31.8\% better MCC values and 7.7--24.1\% higher F1-Score values for the seven metrics than the baseline models without data augmentation.
We also confirmed that the best models using data augmentation were statistically significantly better than the respective baselines in terms of both MCC and F1-Score, based on the non-parametric Wilcoxon signed-rank tests~\mbox{\cite{wilcoxon1992individual}} with $p$-values $<$ 0.01 and non-negligible effect sizes ($r \geq 0.1$), as shown in the last two rows of \tab~\ref{tab:da_results}.
The significant improvements in performance of data augmentation highlight the importance of mitigating the data imbalance issue for the SV assessment tasks.

Regarding the performance of individual CVSS metrics, data augmentation was particularly effective for Access Vector, Access Complexity, Authentication, and Severity.
Notably, the improvement of using data augmentation over the baseline for Access Vector could go up to 81.3\% in MCC and 40.5\% in F1-Score, i.e., using \code{RF + Doc2Vec} with Over-Sampling.
This finding can be explained by the fact that these three metrics had the smallest size of the minority classes (with the fewest samples, as depicted in \fig~\ref{fig:cvss2_classes}.
It is worth noting that Authentication did not always have the highest improvement value despite having the smallest minority class, i.e., Multiple.
This was mostly because the Multiple class did not appear in all the evaluation rounds, and the impacts of data augmentation were also attributed to the Medium class, which was larger than the minority classes of Access Vector, Access Complexity, and Severity.
The Impact metrics (Confidentiality, Integrity, and Availability) also benefited less from data augmentation than the other metrics. This result is likely because these three metrics had the least imbalances in the data classes among the CVSS metrics (see \fig~\ref{fig:cvss2_classes}).

\begin{tcolorbox}
\textbf{RQ1 Summary}. Mitigating data imbalance can have significantly positive impacts on the SV assessment models. Data augmentation improves the baseline predictive performance of all the seven CVSS metrics, with increases of 5.3--31.8\% in MCC and 7.7--24.1\% in F1-Score.
Exploitability and Severity CVSS metrics exhibit more performance gains with data augmentation than the Impact metrics, likely because of the higher degrees of data imbalance.
\end{tcolorbox}

\begin{figure*}[t]
  \begin{subfigure}{\textwidth}
   \centering
   \includegraphics[trim={0.1cm 2.7cm 0.1cm 2.7cm},clip,width=\columnwidth,keepaspectratio]{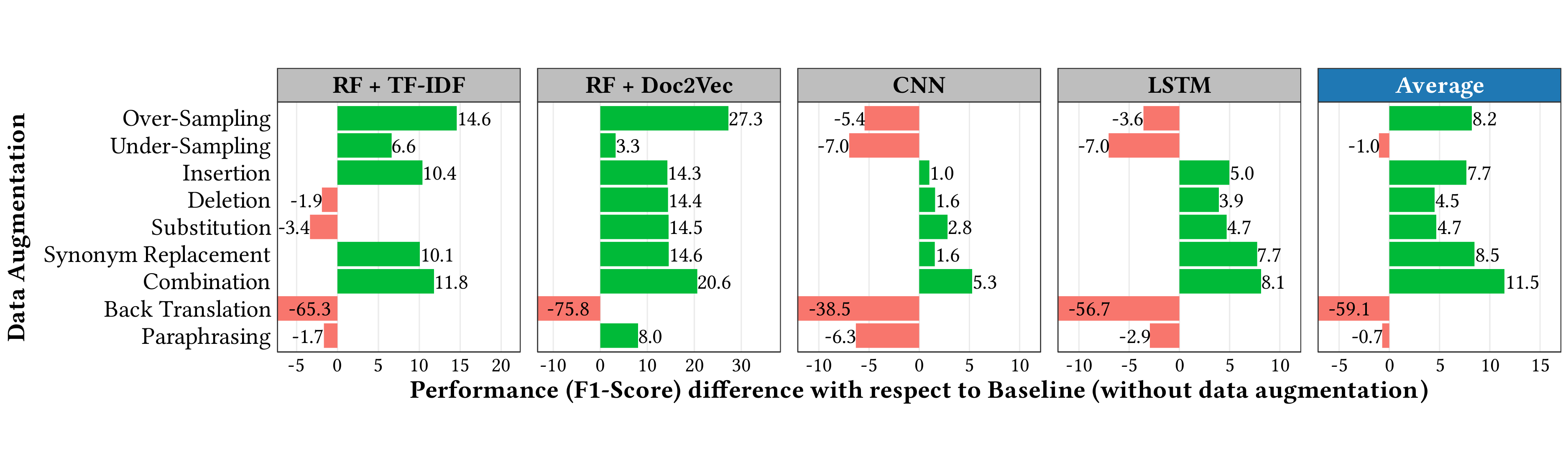}
   \label{fig:f1_diff}
   \end{subfigure}
   \rule{\textwidth}{0.01pt}
   \begin{subfigure}{\textwidth}
   \centering
   \includegraphics[trim={0.1cm 2.7cm 0.1cm 2.7cm},clip,width=\columnwidth,keepaspectratio]{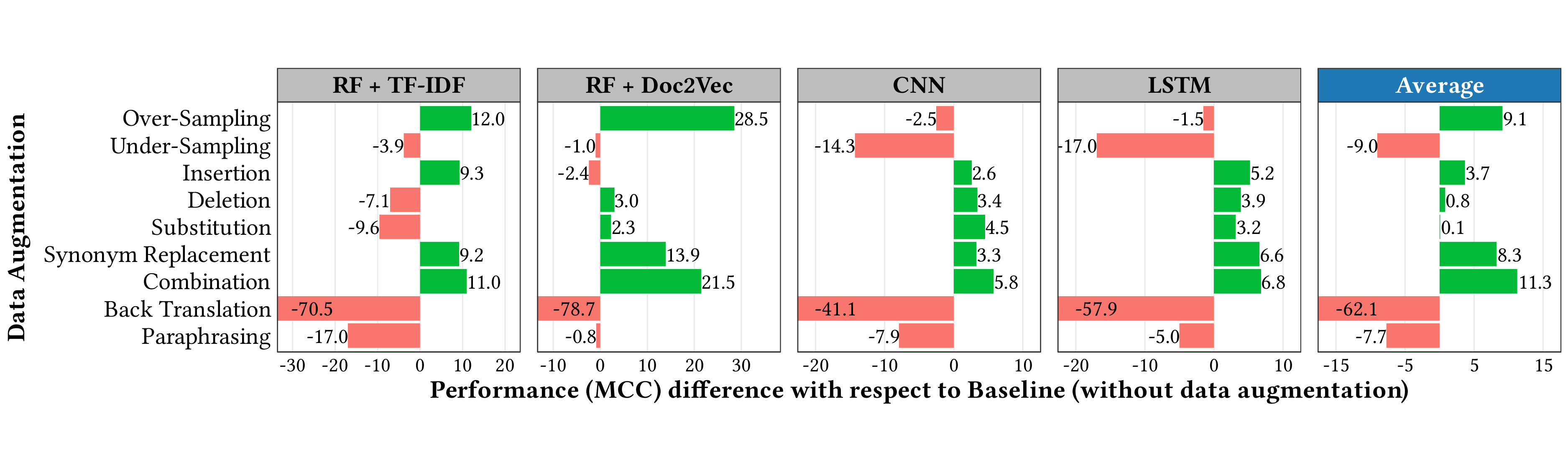}
   \label{fig:mcc_diff}
   \end{subfigure}
   \caption{Percentage (\%) differences in testing SV assessment performance (F1-Score and MCC) between using different data augmentation techniques and the baseline (without data augmentation) across different model types.}
  \label{fig:da_comparisons}
\end{figure*}

\subsection{RQ2: Performance of Individual Data Augmentation Techniques}
\label{subsec:rq2_results}

Expanding upon the overall improvement of data augmentation over the baseline in RQ1, the RQ2 results showed that more than half (6/9) of the studied augmentation techniques were better than the None baseline case (see \fig~\ref{fig:da_comparisons}). On average, the outperforming data augmentation techniques were Simple Text Augmentation (Combination, Synonym Replacement, Insertion, Deletion, and Substitution) and Random Over-Sampling. The performance analysis of individual data augmentation techniques is presented hereafter.

The Combination technique had the highest average performance among the data augmentation techniques across the four studied models. Combination, on average, improved the baseline by 11.3\% in MCC ($p$-value = 3.7e-9, $r$ = 0.788) and 11.5\% in F1-Score ($p$-value = 7.5e-9, $r$ = 0.772), as shown in \tab~\ref{tab:da_results}.
Particularly, \tab~\ref{tab:da_results} shows that Combination was the best data augmentation technique (i.e., the models with the highest MCC values) for 5/7 CVSS metrics, i.e., Access Vector, Authentication, Confidentiality, Integrity, and Availability.
This finding shows that such simple textual modifications that have been successful for other text classification tasks~\cite{wei2019eda} are also helpful for augmenting SV descriptions and improving the performance in classifying the CVSS metrics.

Among the techniques used in Combination, Synonym Replacement was the best-performing operation across different model types, except CNN (see \fig~\ref{fig:da_comparisons}).
This technique was also the best data augmentation technique for Access Complexity and Severity.
In addition, we found that Insertion, together with Synonym Replacement, contributed more to overall performance improvements of Combination, discussed in the previous paragraph, than Deletion and Substitution.
Synonym Replacement and Insertion kept original words unchanged and/or replaced the words with others having similar meanings.
On the other hand, Deletion and Substitution can alter the meaning more by removing (important) information/words in SV descriptions.
Thus, Synonym Replacement and Insertion are more likely to keep the original meaning for augmented descriptions than Deletion and Substitution.
Despite that, Deletion and Substitution still provided higher performance improvements than without using data augmentation.

Random Over-Sampling, though simple, still proved its usefulness for boosting the average SV assessment performance, on par with Synonym Replacement.
This technique was particularly effective for the ML models (i.e., \code{RF + TF-IDF} and \code{RF + Doc2Vec}), yet was not as much for the DL models (i.e., \code{CNN} and \code{LSTM}).
In addition, we discovered that Random Over-Sampling was better than Random Under-Sampling. Given that Under-Sampling has been previously used for SV assessment~\cite{han2017learning}, our finding identifies and emphasizes the sub-optimality of Random Under-Sampling. Instead, we suggest that Random Over-Sampling or Simple Text Augmentation techniques should be used for the SV assessment tasks for improved performance.

Despite incorporating the context of whole SV descriptions, Contextual Text Augmentation (Back Translation and Paraphrasing) could not improve the performance of the SV assessment models.
Through a closer inspection, we found that these two techniques at times could not properly comprehend important and software/SV-specific words in many of the augmented SV descriptions, potentially leading to information loss or semantic changes in the descriptions during model training.
For instance, the word ``\textit{passwd}'' (command to change password) was changed to the general ``\textit{password}'' by both Back Translation and GPT-based Paraphrasing for the description ``\textit{Buffer overflow in passwd in BSD based operating systems 4.3 and earlier allows local users to gain root privileges by specifying a long shell or GECOS field.}'' of CVE-1999-1471. Such change could have obscured the information about the location of the SV.
It is important to note that it is non-trivial to automatically identify such words to preserve. Automatic preservation of security/SV-specific words in SV descriptions may be an interesting direction to explore in the future.

We also discovered that the best data augmentation technique also varied for each of the four studied types of SV assessment models, as shown in \fig~\ref{fig:da_comparisons}.
In terms of MCC averaging across seven SV assessment tasks/metrics, \code{RF + TF-IDF} and \code{RF + Doc2Vec} achieved the best improvements of 10.1\% and 25.1\% with Random Over-Sampling, compared to the optimal improvements of 5.5\% for \code{CNN} and 6.8\% for \code{LSTM} when combined with the Combination technique.
The MCC performance gains of data augmentation for \code{RF + TF-IDF}, \code{RF + Doc2Vec}, \code{CNN}, and \code{LSTM} have been confirmed statistically significant using the non-parametric Wilcoxon signed-rank tests~\mbox{\cite{wilcoxon1992individual}} with $p$-values of 1e-4, 2.2e-12, 9.5e-4, 1.4e-5 and medium/large effect sizes of 0.471, 1.36, 0.341, and 0.388, respectively.
We also observed similar improvements in terms of F1-Score for the four model types when coupled with the same respective data augmentation techniques.
The discrepancies in performance gains between ML and DL can be attributed to the different nature of feature representation and learning of these two model types.
The DL-based models partially (\code{CNN}) or fully (\code{LSTM}) extracted the word sequence of SV descriptions, which can better capture the structure and semantics of the descriptions than the ML models, e.g., changing the word order would affect the DL models more than the ML ones.
Thus, the performance of DL would be more sensitive to the textual changes, e.g., by the Combination and Synonym Replacement data augmentation techniques, compared to just repeating the same features multiple times by Random Over-Sampling.

\begin{tcolorbox}
\textbf{RQ2 Summary}.
Among the studied Data Augmentation (DA) techniques, Combination (randomly inserting, deleting, and replacing text) performs the best with an average MCC improvement of 11.3\% over the baseline.
Synonym Replacement, Random Over-Sampling, and Insertion are also substantially better than the baseline.
Random Under-Sampling and Contextual Text Augmentation, i.e., Back Translation and Paraphrasing, are worse than the baseline, probably due to missing information and/or lacking semantic understanding of software/SV-specific words.

\textbf{\textit{Task-wise}}. The Combination technique is the best for Access Vector, Authentication, Confidentiality, Integrity, and Availability. Synonym Replacement is the optimal technique for the Access Complexity and Severity.

\textbf{\textit{Model-wise}}. The best DA techniques for DL and ML are Combination and Random Over-Sampling, respectively, showing that DL benefits more from textual changes than ML.
These techniques improve the MCC values of the Machine Learning (ML) and Deep Learning (DL) models by 10.1--25.1\% and 5.5--6.8\%, respectively.

\end{tcolorbox}

\section{Discussion}
\label{sec:discussion}

\subsection{Advancing Data-Driven SV Assessment: Data Augmentation and Beyond}\label{subsec:implications}

In recent years, data-driven approaches have been increasingly used for SV assessment. One of the main goals in the field is to increase the performance of developed models~\cite{le2022survey}.
Our study findings in Section~\ref{sec:results} have highlighted data augmentation as an effective approach to achieving this goal for all of the CVSS tasks.
Furthermore, the studied data augmentation techniques work directly on the input data and independently of underlying models.
Thus, they can be seamlessly integrated with and potentially enhance the performance of almost any SV assessment models without changing the model architectures, ranging from the existing well-known models (RF, CNN, and LSTM) to newly proposed models in the future.

We recommend that Combination (i.e., combining Word Insertion, Deletion, and Substitution/Replacement) should be considered as a baseline of data augmentation for SV assessment in the future because this technique has been shown to improve the performance across the board.
The demonstrated effectiveness suggests that Combination-augmented descriptions, despite having textual changes, can still retain the semantics/label of the respective original descriptions.
Following the prior studies~\cite{tian2012identifying,wei2019eda,le2020puminer}, we set out to validate this conjecture by comparing the cosine similarities between the feature vectors of the augmented descriptions to the centroids (average feature vectors of the original descriptions) of each class CVSS-metric-wise.
For each metric, the features were extracted from the optimal model trained with the Combination data augmentation technique.
\tab~\ref{tab:sim_analysis} shows that the augmented descriptions were indeed more similar to those of the original class than the other classes for all the metrics.
Such results can increase the confidence in using this data augmentation technique for SV assessment as it mainly makes structural rather than semantic changes, i.e., often retaining the original label.

\begin{table}[t]
\caption{Average cosine similarities between the Combination-augmented descriptions and the original descriptions of the same and other classes for each of the CVSS metrics. \textbf{Note}: The \textit{Other} cells are the maximum values among the other classes. Higher value is better.}
\label{tab:sim_analysis}
\resizebox{\columnwidth}{!}{%
\begin{tabular}{l|ccccccc|c}
\hline
\multirow{2}{*}{\textbf{Sim.}} & \multicolumn{7}{c|}{\textbf{CVSS Metrics}}                                                  & \multirow{2}{*}{\textbf{Avg.}} \\ \cline{2-8}
                              & \textbf{AV} & \textbf{AC} & \textbf{Au} & \textbf{C} & \textbf{I} & \textbf{A} & \textbf{S} &                                \\ \hline
\textbf{Same}  & \cellcolor[HTML]{C0C0C0} \textbf{0.223} & \cellcolor[HTML]{C0C0C0} \textbf{0.201} & \cellcolor[HTML]{C0C0C0} \textbf{0.171} & \cellcolor[HTML]{C0C0C0} \textbf{0.288} & \cellcolor[HTML]{C0C0C0} \textbf{0.305} & \cellcolor[HTML]{C0C0C0} \textbf{0.420} & \cellcolor[HTML]{C0C0C0} \textbf{0.262} & \cellcolor[HTML]{C0C0C0} \textbf{0.267} \\ \hline
\textbf{Other} & 0.208 & 0.184 & 0.151 & 0.275 & 0.293 & 0.408 & 0.250 & 0.252 \\ \hline
\end{tabular}%
}
\end{table}

Despite the success of data augmentation techniques for SV assessment, there is still room for improvement for these techniques and SV assessment as well. We analyzed 3,668 SVs in the testing sets where the optimal models (with the highest testing MCC values) trained with and without augmented SV descriptions could not correctly predict for all of the seven CVSS metrics. These cases showed the scenarios where data augmentation consistently struggled to provide meaningful improvement to CVSS assessment models.
From the analysis, we identified a common pattern of these incorrect cases.
Data augmentation had difficulty in improving the assessment performance when the input SV description was short/uninformative.
The average number of words in these problematic descriptions was only 16 compared to 28 in all SV descriptions in the testing sets.
An example was ``\textit{static/js/pad\_utils.js in Etherpad Lite before v1.6.3 has XSS via window.location.href.}'' -- the description of CVE-2018-6834.
We posit that these cases lack the information/words about some characteristics of SVs, e.g., SV impact in the presented example. Moreover, many of the words were software-specific terms such as ``\code{static/js/pad\_utils.js}'' or ``\code{window.location.href}''. 
If the words of such descriptions were randomly removed or replaced, which are the key operations of the Combination data augmentation technique, the information loss would be further increased.
Such words also did not appear in the WordNet, making Synonym Replacement struggle to find a suitable synonym.
In addition, such keywords could not (yet) be effectively comprehended by Contextual data augmentation, i.e., Back Translation or Paraphrasing.
In the future, multiple sources such as social media sites and/or external security advisories can be leveraged to provide more informative descriptions of such SVs. Still, more research is required to automatically validate the relevance and trustworthiness of the externally gathered information~\cite{anwar2021cleaning}.

\subsection{Threats to Validity}\label{subsec:threats_to_validity}

\noindent \textbf{Internal validity}.
A possible threat here is that our optimal models may not guarantee the highest performance for SV assessment. However, we assert that it is nearly impossible to achieve this because there are infinite values of hyperparameters of the models to tune.
Our study may not provide the best possible results for SV assessment; however, it still highlights the benefits of using data augmentation for handling the data imbalance issue of the tasks and provides the baseline performance of SV assessment with data augmentation for future research to build upon.

\noindent \textbf{External validity}.
Our work may not generalize to all SVs. We tried to mitigate the threat by using NVD -- one of the most comprehensive repositories of SVs. Our dataset contained more than 180k SVs, ranging from 1988 to 2023.
There is also a potential concern about the generalizability of our findings to other SV assessment models. We mitigated this threat by investigating the four most commonly used baseline models for SV assessment, which are expected to provide direct contributions to the current practices of the field.
We also release our data, code, and models at~\cite{reproduction_package_esem2024} for future research to replicate our study on new SV data and models.

\noindent \textbf{Conclusion validity}.
We mitigated the randomness of the results by taking the average value of multi-round time-based evaluation. The key performance comparisons of different SV assessment models with and without data augmentation were also confirmed using the non-parametric Wilcoxon signed-ranked tests with $p$-values $<$ 0.01 and non-negligible values of the effect size.

\section{Related Work}
\label{sec:related_work}

\subsection{Data-Driven SV Assessment and Analysis}
\label{subsec:data_driven_sap}

SV assessment is a crucial process in dealing with SVs, and CVSS offers one of the most dependable metrics for such assessment~\cite{johnson2016can}. Prior research delved into analyzing CVSS metrics and SV trends by integrating diverse SV data from multiple SV repositories \cite{murtaza2016mining,almukaynizi2019patch}, security advisories \cite{edkrantz2015predicting,huang2013novel}, dark web sources \cite{almukaynizi2019patch,nunes2016darknet}, and social networks like Twitter/X \cite{sabottke2015vulnerability}. However, these studies operated under the assumption that all CVSS metrics were available during the analysis, which has been shown to be unrealistic in real-world settings~\cite{gong2019joint}. In contrast, our work deviates from this assumption by utilizing solely the SV descriptions, making our method more adaptable and suitable for both new and older SVs.

Bozorgi et al.~\cite{bozorgi2010beyond} were among the first to employ data-driven models for SV assessment utilizing solely SV descriptions as inputs. The authors used an SVM model and several attributes such as NVD description, CVSS, and publication dates, to gauge the likelihood of exploitation and time-to-exploit of SVs. This pioneering work sparked a substantial volume of subsequent research aimed at automating SV assessment tasks using data-driven models~\cite{le2022survey}.
Numerous recent studies (e.g.,~\cite{yamamoto2015text,spanos2017assessment,spanos2018multi,le2019automated,elbaz2020fighting,duan2021automated}) have drawn on SV descriptions found in bug/SV reports/databases, mainly NVD, to predict the CVSS metrics for ever-increasing SVs.

Although these studies have demonstrated the promising use of ML/DL for SV assessment, they hardly addressed the inherent data imbalance problem of the tasks. To the best of our knowledge, Han et al.~\cite{han2017learning} were the only prior study that attempted to tackle the problem using Random Under-Sampling. However, our results in Section~\ref{sec:results} have shown that Random Under-Sampling was sub-optimal for SV assessment and could even reduce the performance of the prediction models. Furthermore, our study is the first to show the potential of other data augmentation techniques, such as combining text insertion, deletion, and substitution/replacement, to effectively mitigate data imbalance for SV assessment.

\subsection{Data Augmentation for Data-Driven Software Engineering Tasks}

Data augmentation has witnessed growing success in the Natural Language Processing domain in recent years~\cite{feng2021survey}. Such success has then inspired the increasing use of data augmentation in the Software Engineering (SE) domain, mainly because many software artifacts are in the form of text~\cite{zhuo2023data}. So far, data augmentation has been applied to a wide range of automated SE tasks such as code clone detection~\cite{pinku2023pathways,liu2023contrabert}, defect prediction~\cite{allamanis2021self}, code summarization~\cite{zhang2020retrieval}, and code question answering~\cite{huang2021cosqa,park2023contrastive}.
These studies have revealed the advantages of augmented data to address the data imbalance/scarcity and overfitting issues of the respective tasks.
However, there is much less work on utilizing data augmentation for software security, particularly SV assessment and analysis -- an integral step in secure software development.
Our work aims to contribute to the body of knowledge in this emerging area by showing the possible benefits and use of data augmentation for SV assessment.
Our promising results in Section~\ref{sec:results} can inspire future work to investigate more sophisticated data augmentation techniques for SV assessment tasks.
While current approaches mainly leverage SV reports for SV assessment as shown in Section~\ref{subsec:data_driven_sap}, future research can also explore code-based data augmentation techniques to complement the text-based techniques investigated in this study with source code for predicting SV assessment metrics.

\section{Conclusion}
\label{sec:conclusions}

We highlighted the importance of mitigating data imbalance for SV assessment.
We investigated the effectiveness of addressing the issue for different SV assessment tasks using nine data augmentation techniques.
Through extensive experiments on 180k+ real-world SVs, we showed that data augmentation could improve the performance of the models without data augmentation by up to 31.8\% in MCC and 24.1\% in F1-Score, particularly the Exploitability and Severity CVSS metrics.
Among the data augmentation techniques, we found that combining simple textual operations, including random text insertion, deletion, and substitution/replacement, achieved the best performance improvements over the baseline.
Our study encourages further investigations into better data augmentation for SV assessment, particularly the techniques that can comprehend software/SV-specific words in SV descriptions.

\section*{Acknowledgments}
The work has been supported by the Cyber Security Research Centre Limited whose activities are partially funded by the Australian Government's Cooperative Research Centres Program. We would like to acknowledge Tuan Lu Dinh for his technical support at the early stage of this work.

\balance

\bibliographystyle{ACM-Reference-Format}
\bibliography{reference}

\end{document}